# Role of structural defects on exchange bias in the epitaxial CoO/Co system

M. R . Ghadimi*, B. Beschoten and G. Güntherodt

II. Physikalisches Institut, RWTH Aachen, 52056 Aachen, Germany

We have studied the influence of non-magnetic defects throughout the antiferromagnet $Co_{1-y}O$ on the exchange bias (EB) in epitaxially grown $Co_{1-y}O$/Co bilayers. These defects are either substitutional or structural (twin boundaries and surface morphology) which both lead to an increase of the EB field. We find a dominance of twin boundaries over surface morphology (roughness) in enhancing EB which is consistent with the domain state model for exchange bias. In contrast, the crystal orientation of the $Co_{1-y}O$ layer does not show a significant effect on the EB in this system.

Exchange coupling at the interface between an antiferromagnet (AFM) and a ferromagnet (FM) causes unidirectional anisotropy in the FM layer, which results in a shift of the hysteresis loop along the magnetic field axis, the so-called exchange bias shift [1,2]. In order to understand the microscopic origin of EB, the domain state (DS) model was proposed, which has been supported by Monte Carlo simulations [3,4]. This model is based on the physics of diluted antiferromagnets in an external magnetic field (DAFF) and yielded the description of the most salient EB features of any model to date [3]. It has been shown that non-magnetic defects in the bulk of the AFM stabilize a domain state after field-cooling below its Néel temperature ($T_N$). These domains carry a remanent domain state magnetization ($M_{DS}$). Their irreversible part $M_{IDS}$ provides the EB at the interface to the FM layer. The number and size of the domains in the AFM can be controlled by the number of defects [5,6].

In our previous experimental studies using twinned $Co_{1-y}O(111)/Co$ bilayers we have shown that the EB field ($B_{EB}$) can be controlled and increased by intentionally diluting the volume part of $Co_{1-y}O$ by non-magnetic defects [5,6]. The DS model predicts that the EB vanishes in the undiluted limit ($Co_{1-y}O$ with y→0). However, experimentally we have still observed a small EB shift at low temperatures even in nominally undiluted samples, which we attributed to residual disorder other than substitutional defects.

In this paper, we present experimental evidence of the direct influence of structural defects on the EB in $Co_{1-y}O$, such as twinning and surface morphology, which we distinguish from the effect by substitutional defects. We show that these additional defects have to be taken into account for the complete assessment of the EB. Additionally, we show that the crystal orientation of the CoO layer does not play an important role for EB in this system.

We have studied AFM/FM bilayers of the CoO(111)/Co(111) system grown on MgO(111) substrates by molecular beam epitaxy. At first, this model system allows to introduce controlled substitutional point defects at the Co sites of the AFM by changing the partial pressure of the oxygen $p(O_2)$ during the growth of the CoO layer. The over-oxidation of CoO under high pressure $p(O_2)$ yields a $Co^{2+}$-deficient layer denoted as $Co_{1-y}O$. Secondly, we can change the stacking order of the bilayer system. For stacking type I (MgO/$Co_{1-y}O$/Co) we obtain untwinned CoO layers. In contrast, for stacking type II (MgO/Co/$Co_{1-y}O$) we observe twinning in the CoO layer which results from the grain boundaries. Hence, we can examine the influence of structural defects on EB by controlling the twin boundaries in $Co_{1-y}O$. All samples have a layer thickness of 20 nm for $Co_{1-y}O$ and 6 nm for Co.

The structural quality and defects have been characterized by *in situ* reflection high-energy electron diffraction (RHEED). A series of the RHEED patterns of both untwinned and twinned $Co_{1-y}O$ and of the respective Co layers for samples of type I and II is depicted in Fig. 1 and its insets. Figures 1(a) and 1(b) show the RHEED patterns of an untwinned $Co_{1-y}O$ layer grown at $4\times10^{-7}$ mbar (undiluted) and $5\times10^{-6}$ mbar (optimally diluted). The electron beam direction is parallel to the direction [1-21] (0°, left panel) of the (111)-oriented MgO substrate and parallel to the [0-11]-direction (30°, right panel). The RHEED patterns of the corresponding twinned $Co_{1-y}O$ layers without and with intentional dilutions are depicted in Fig. 1(c) and 1(d), respectively. All RHEED patterns of the $Co_{1-y}O$ layers (Fig. 1(a)-(d)) exhibit spots which are not aligned on Laue circles indicating a three dimensional (3D) growth in both stacking orders. For diluted $Co_{1-y}O$ (Figs. 1(b) and 1(d)) grown at $5\times10^{-6}$ mbar, the destructive interference from the fcc lattice is suppressed due to some empty lattice sites which are created by the substitutional defects. Hence, additional diffraction spots become visible, which correspond to a crystalline structure with approximately twice the lattice constant in real space (compare Fig. 1(a) with 1(b) for 0°). We now discuss the RHEED patterns along the 30° direction. Only for stacking type II we observe additional spots in the RHEED patterns of the $Co_{1-y}O$(111) layer (Figs. 1(c) and 1(d)) which we attribute to the twinning, where crystallites are rotated by 60° relative to each other. A more detailed analysis of the twin structure is given in Refs. [6,7]. From the RHEED patterns we deduced values of 0.42 nm and 0.83 nm for the lattice parameters of undiluted and diluted $Co_{1-y}O(111)$, respectively.

For all samples the Co layer (insets of Fig. 1(a)-(d)) grows in a (111)-oriented fcc lattice structure with a lattice parameter of 0.35 nm as deduced from the RHEED. While the Co layer grows as a "smoothened" 3D layer on MgO (see streaky spots in the insets of Fig. 1(c) and 1(d)), the $Co_{1-y}O$ layer directly grown on MgO shows a typical 3D-growth (Fig. 1(a) and 1(b)). This difference in surface roughness has to be compared with the effect of twinning to assess the role of different types of defects on EB (see below). The epitaxial relationships obtained from the RHEED patterns for 0° or 30° between the MgO(111) substrate as well as the twinned and untwinned bilayers, respectively, are

$(111)MgO \parallel (111)Co_{1-y}O \parallel (111)Co$; $[0-11]MgO \parallel [0-11]Co_{1-y}O \parallel [0-11]Co$ and

$(111)MgO \parallel (111)Co \parallel (111)Co_{1-y}O$; $[0-11]MgO \parallel [0-11]Co \parallel [0-11]Co_{1-y}O$. These results are in agreement with simulated diffraction patterns by means of the CaRIne program and were confirmed by additional x-ray diffraction measurements (not shown).

The magnetic hysteresis loops are measured by a Quantum Design SQUID (superconducting quantum interference device) magnetometer. The samples are cooled in the presence of an external magnetic field $B_{cool} = 1$ T from 320 K through the Néel temperature $T_N = 291$ K of CoO to 5 K with the external field oriented parallel to the plane of the CoO film along the [1-21] direction. To eliminate artefacts related to the training effect [6], the sample was field-cycled several times prior to the measurement. The exchange bias field $B_{EB}$ is determined from the shift of the magnetic hysteresis loop.

The dilution dependence of the EB field $|B_{EB}|$ is shown in Fig. 2(a) for both stacking types with twinned and untwinned $Co_{1-y}O$ layers. We observe a strong dependence of $|B_{EB}|$ on the oxygen partial pressure $p(O_2)$, i.e. on the dilution $y$ of $Co_{1-y}O$, exhibiting a maximum near $4.5 \times 10^{-6}$ mbar. For both twinned as well as untwinned $Co_{1-y}O$ the $|B_{EB}|$ first increases with increasing dilution, reaching a maximum and then decreases as $p(O_2)$ rises further. Note that the EB shift for the samples with twinned $Co_{1-y}O$ is larger at almost all dilutions compared to the samples with untwinned $Co_{1-y}O$, except for the sample prepared at the highest oxygen pressure. This additional EB shift indicates that structural defects in $Co_{1-y}O$, such as twin boundaries can enhance the EB effect independent of the concentration of additional substitutional defects.

A longstanding issue in the theory of EB is the role of roughness at the interface between FM and AFM layers as first pointed out by Malozemoff [8]. Based on the RHEED patterns (insets in Fig. 1(c) and 1(d)) it become apparent that the Co layer grows as a "smoothened" 3D-layer on MgO (with twinned $Co_{1-y}O$ growing on top), while both the (untwinned) $Co_{1-y}O$ (Fig. 1(a) and 1(b)) and the Co layer on top show a 3D-growth. Although the interface roughness in the twinned samples is less pronounced than in the untwinned samples, the resulting EB shift is larger (see Fig. 2(a)). This provides experimental evidence for the dominance of twin boundaries in the bulk of the AFM layer over the relevance of interface roughness for EB which is consistent with the DS model.

The implementation of non-magnetic defects in the AFM by varying $p(O_2)$ fulfills two important goals. On the one hand, these non-magnetic defects support the formation of domains in the AFM. On the other hand, they create uncompensated spins in the AFM, giving rise to the irreversible domain state magnetization $M_{IDS}$, which is aligned by the external field during field cooling. The part of the $M_{IDS}$ at the interface to the FM layer is responsible for EB. In our EB system, the increasing density of non-magnetic substitutional defects with increasing $p(O_2)$ favors the development of domains in the AFM. The domain walls pass preferentially through these defects to minimize the energy of the system. This process leads to an increase of $M_{IDS}$, which at the AFM/FM interface gives rise to an increase of $|B_{EB}|$. The maximum of $|B_{EB}|$ near $5\times10^{-6}$ mbar (Fig. 2(a)) corresponds to optimal dilution and thus to a maximum of $M_{IDS}$. For larger dilution ($p(O_2)$ higher than $5\times10^{-6}$ mbar) the AFM spin lattice looses its connectivity between the spins, i.e. the AFM order is increasingly reduced. This is most pronounced at $p(O_2)=2\times10^{-5}$ mbar, for which the higher defect concentration in the twinned sample leads to a stronger decrease of $|B_{EB}|$ compared to the untwinned sample. In other words, this leads to the creation of magnetically disconnected spin clusters, which do not contribute to a net magnetization in the AFM, i.e. $M_{IDS}$ decreases. As a consequence $|B_{EB}|$ decreases.

Furthermore we observe an enhancement of EB with additional structural defects. Like for the substitutional defects the domain wall energy will also be reduced when passing through a twin boundary thus favoring the formation of additional domains in the AFM. Hence, these structural defects might also lead to an increase of $M_{IDS}$ and thus enhance $|B_{EB}|$ as seen in Fig. 2(a).

However, even without substitutional defects and after eliminating the twin boundaries there is still a remaining EB shift of about 12 mT for the sample with CoO prepared at $p(O_2) = 4\times10^{-7}$ mbar (Fig. 2(a)). The origin of this residual EB shift might be explained by the presence of further imperfections in the CoO, caused by the 3D surface morphology as observed by RHEED. In our case, the pronounced 3D surface morphology of the $Co_{1-y}O$ films, which is due to the high energy of the (111)-oriented surface of fcc transition metal-oxides [9] (Fig. 1(a)-(d)) can favour the development of a higher density of domain walls in the AFM. The study of different morphological properties of the untwinned-undiluted CoO (111) layers by varying the growth parameters is not feasible, because the surface morphology of these polar CoO(111) surfaces is mostly controlled by thermodynamics. Therefore, we additionally prepared (100)-oriented type I CoO/Co samples which are untwinned and undiluted. The nonpolar CoO (100) surface can be tailored to change from a "smoothened" 3D to a 3D morphology by changing the growth parameters.

Figure 2(b) shows the temperature dependence of $|B_{EB}|$ for two (untwinned and undiluted) MgO(100)/CoO(100)/Co(11-20)/Au bilayers with a 3D- and a "smoothened" 3D-growth of the CoO. Both samples are prepared under the same conditions ($p(O_2) = 4\times 10^{-7}$ mbar), but for different degassing temperatures of the MgO substrate. For a degassing temperature of 800 °C we obtain "smoothened" 3D morphology, whereas for 1100 °C a 3D surface morphology is obtained. The difference in the surface morphology of the CoO layers is depicted by the RHEED patterns in the insets of Fig. 2(b). The sample with the "smoothened" 3D-surface of the CoO layer has a significant lower EB shift by about a factor of 3 up to 200K. This result supports the notion that the 3D surface morphology of the CoO layer creates additional defect that can enhance the EB. Note that a systematic investigation of the structural defects on EB in (100)-oriented $Co_{1-y}O$/Co was not possible due to difficulties in preparing twinned $Co_{1-y}O$(100) layers on MgO(100). In previous experiments by Stöhr et al. [10] similar correlation between structural effects and AFM domains was found.

We emphasize that the influence of crystal orientation on EB does not seem to play an important role in our EB system. We only find slightly larger values of the EB shift in the case of untwinned-undiluted (100)-oriented films ($|B_{EB}|_{(100)}$ = 17 ± 0.5 mT at 5 K) rather than untwinned-undiluted (111)-oriented films ($|B_{EB}|_{(111)}$= 12 ± 0.5 mT at 5 K). This result clearly disproves the statement that EB results only from an uncompensated surface of (111)-oriented CoO [11]. In addition, the $|B_{EB}|$ of the "smoothened" 3D CoO layer decreases nearly linearly up to the blocking temperature, while in the case of the 3D-growth it decreases weakly between 5K and 230K. Only above 230K the $|B_{EB}|$ drops rather strongly. This can be interpreted as a fingerprint of increased domain wall pinning at morphological defects, such as dislocations, grain boundaries, islands edges and ridges in the CoO film with a 3D-growth mode.

In conclusion, we have shown experimentally that any kind of disorder in the volume part of the AFM, such as substitutional defects (dilution) or structural defects (twinning or surface morphology), significantly enhances EB. These defects support the formation of volume domains in the AFM which triggers the EB at the AFM/FM interface. The stronger increase of $|B_{EB}|$ by twin boundaries which act in the volume of the AFM as compared to surface roughness is consistent with the domain state model for exchange bias. In addition, we have demonstrated that the crystal orientation does not seem to play an important role in epitaxially grown CoO/Co.

Work partially supported by the DFG/SPP1133 and the European Community's Human Potential Program/NEXBIAS.

* Email address: ghadimi@physik.rwth-aachen.de

FIGURE 1

RHEED images of 20 nm thick $Co_{1-y}O$ as (a) untwinned-undiluted CoO on MgO(111), (b) untwinned-diluted CoO on MgO(111) as first layer, (c) twinned-undiluted CoO on Co(111)/MgO(111) and (d) twinned-diluted CoO on Co(111)/MgO(111). The two vertical panels show the patterns for 0° (left panel) and 30° (right panel) in-plane orientation of the incident electron beam relative to the MgO(111). In addition to the patterns of the CoO layers, the insets show the RHEED images of the 6 nm thick fcc Co for (a)(b) as second layer on CoO and for (c)(d) as first layer on MgO.

FIGURE 2

(a) EB field at 5K as function of the oxygen pressure during deposition of the $Co_{1-y}O$ layer for twinned and untwinned AFM in $Co_{1-y}O(111)/Co(111)$ on MgO(111).

(b) Temperature dependence of the EB field for "smoothened" 3D- and 3D-growth of untwinned-undiluted CoO in MgO(100)/CoO(100)/Co(11-20). Additionally, the RHEED patterns of the CoO layers are depicted. The electron beam direction is parallel to the [001] direction of the (100)-oriented MgO substrate.

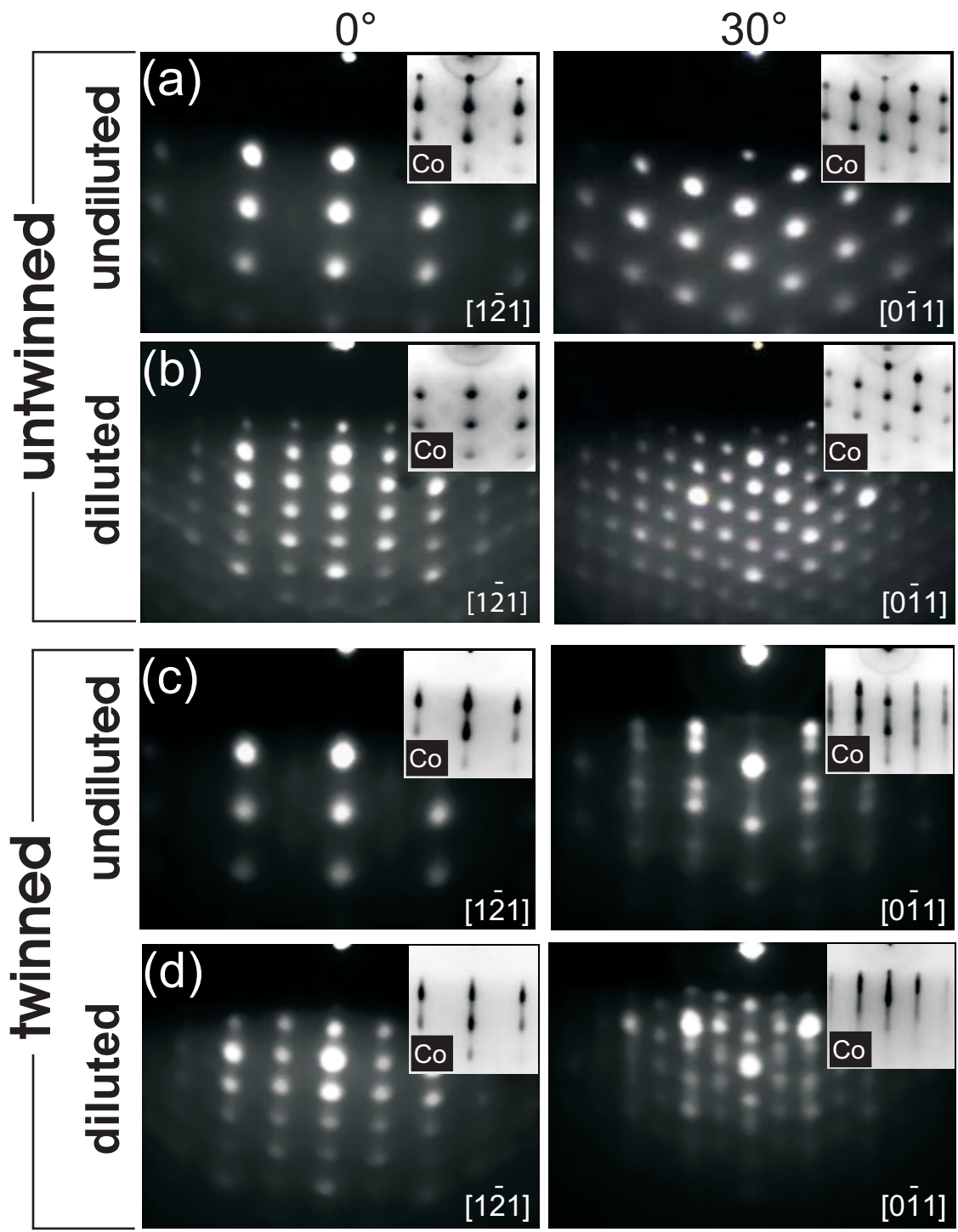

Figure 1: M. R. Ghadimi et al.

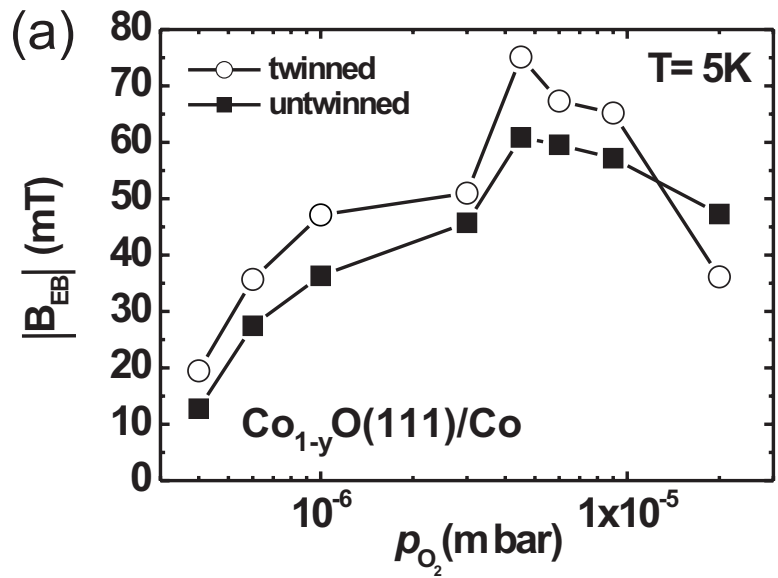
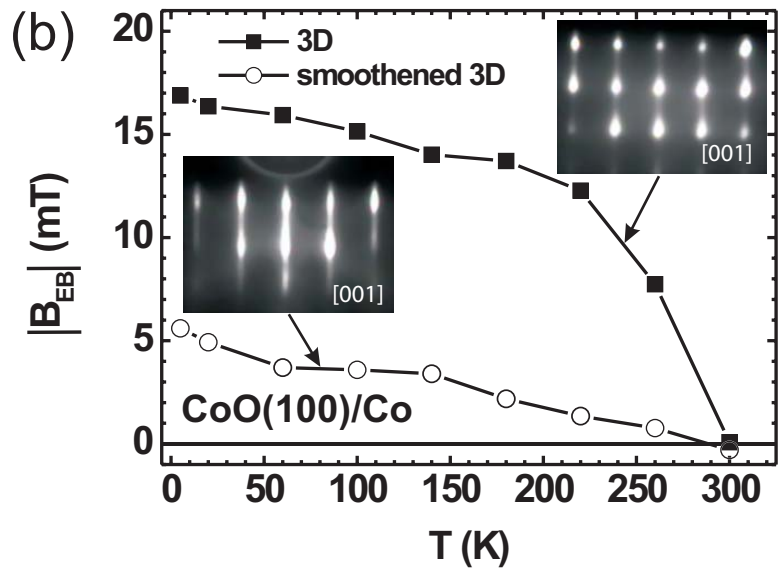

Figure 2: M. R. Ghadimi et al.